\documentclass[twocolumn,showpacs,preprintnumbers,amsmath,amssymb,prl]{revtex4}

\usepackage{graphicx}
\usepackage{dcolumn}

\usepackage{mathptmx, courier, pifont}
\usepackage[scaled=0.92]{helvet}
\usepackage[T1]{fontenc}
\usepackage{textcomp}

\usepackage{bm}

\begin{document}

\title{Nonaxial-octupole effect in superheavy nuclei}

\author{Y.-S. Chen$^{1}$, Yang Sun$^{2,3}$, Zao-Chun Gao$^{1,4}$}

\affiliation{ $^{1}$China Institute of Atomic Energy, P.O. Box
275(18), Beijing 102413, People's Republic of China \\
$^{2}$Department of Physics, Shanghai Jiao Tong University,
Shanghai 200240, People's Republic of China \\
$^{3}$Joint Institute for Nuclear Astrophysics, University of
Notre Dame, Notre Dame, Indiana 46556, USA \\
$^{4}$Department of Physics, Central Michigan University, Mount
Pleasant, Michigan 48859, USA}

\date{\today}

\begin{abstract}
The triaxial-octupole Y$_{32}$ correlation in atomic nuclei has
long been expected to exist but experimental evidence has not been
clear. We find, in order to explain the very low-lying 2$^-$ bands
in the transfermium mass region, that this exotic effect may
manifest itself in superheavy elements. Favorable conditions for
producing triaxial-octupole correlations are shown to be present
in the deformed single-particle spectrum, which is further
supported by quantitative Reflection Asymmetric Shell Model
calculations. It is predicted that the strong nonaxial-octupole
effect may persist up to the element 108. Our result thus
represents the first concrete example of spontaneous breaking of
both axial and reflection symmetries in the heaviest nuclear
systems.
\end{abstract}


\maketitle

The issue that underlies the present investigation is whether some
heaviest atomic nuclei can be characterized as having an intrinsic
mean field with nonaxial-octupole component, in connection with
the tetrahedral symmetry which is quite common in molecular,
metallic clusters, and some other quantum objects. The tetrahedral
symmetry is a direct consequence of the point group and
corresponds to the invariance under transformation of the group
T$^D_d$ which has two one- and one four-dimensional irreducible
representation.  In molecular and metallic clusters, the
tetrahedral symmetry is determined by mutual geometric arrangement
of the constituent ions. In atomic nuclei, which are finite
many-body systems with the strong interaction, the situation is
more complex. However, it is usually understood that nuclear
shapes are governed by the shell effects.The tetrahedral
instability in finite many-fermion systems was suggested in 1991
\cite{hamamoto1991}. Tetrahedral shape in nuclei was predicted
with more realistic model calculations \cite{Takami1998,
Dudek2002, Schunck2004, dudek2006}.

For description of nuclear shape, it has been proven to be
extremely useful to parameterize the surface $R(\theta, \phi)$
through its expansion in the spherical harmonic function
$Y^*_{\lambda\mu}(\theta, \phi)$ , where $\theta$ and $\phi$ are
the rotation angles \cite{bmbook}. The coefficients of the
expansion , $\alpha_{\lambda\mu}$, are related to deformation
parameters. The quadrupole deformation ($\lambda$ = 2) is the
leading deformation effect known almost for all the nuclei
throughout the nuclear chart. The next one, the octupole
deformation ($\lambda$ = 3), is related to a spontaneous breaking
of reflection symmetry, or parity, and is known in some nuclei.
The tetrahedral symmetry is realized at the first order through
the nonaxial-octupole Y$_{32}$ deformation
\cite{dudek2003,Zberecki2006}. Almost pure Y$_{32}$ shape was
suggested by the two level model study of finite Fermion
systems\cite{frisk1994}. Calculations of possible nonaxial
pear-like ground-state deformations in the mass region $220 < A <
230$ has been reported \cite{chasman1986}. Both axially- and
nonaxially-enhanced octupole correlations have been suggested for
the $A = 90$ mass region \cite{Butler1996} and in the
superdeformed Hg nuclei \cite{Naka1996}. The Skyrme HFB
calculation has predicted that some nuclei can be extremely soft
against the Y$_{32}$ tetrahedral deformation
\cite{Yamagami2001,olbratowski06}.

One may thus conclude that there have been many theoretical
discussions about this exotic type of nuclear deformation; however,
experimental evidence has not been very clear. This may imply that
the tetrahedral symmetry and the pure Y$_{32}$ shape manifest
themselves in nuclear physics in a very complex way. The octupole
mode, as a smaller deformation effect, has to compete with the
dominant quadrupole mode, and as a result, the appearance of the
tetrahedral symmetry in all known mass regions is somewhat obscure.
The question is whether one can find evidences in the heaviest mass
region which is still less known. It is legitimate to ask this
question because a strong octupole effect has been well known to
exist in the actinide region \cite{Ahmad1993} which is the closest
neighboring mass region to the heaviest nuclei. The present study
will try to answer this question. Some experimental data, though
very limited and incomplete, have already indicated such a
possibility. These data show that very low-lying negative parity
bands with the sequence of $I^\pi=2^-, 3^-, 4^-, 5^-, \cdots$ appear
in the spectrum of some very heavy nuclei (see discussions below).

Exploring the predicted stable island of superheavy elements is the
current goal in nuclear science. In the past few years, progress has
been made in synthesis of new elements (for review, see Refs.
\cite{Hofmann00,Armbruster00,Hofmann01,Oganessian07}). On the other
hand, the heaviest nuclei for which detailed spectroscopic
measurements can be performed are the isotopes with proton number
$Z\sim 100$ and neutron number $N\sim 150$
\cite{Leino04,Rodi04,Green07}. The significance of studying the
deformed nuclei in this transfermium region is that it can provide
an indirect way to access the single-particle states of the next
spherical shells because some of these orbitals are strongly
down-sloping, and thus can come close to the Fermi surface in the
deformed region \cite{Chas77}.

There has been experimental observation of low-energy $2^-$ bands in
the mass-250 nuclei. For example, for the $N=150$ isotonic chain,
early data showed a $2^-$ band in $^{246}$Cm with an bandhead energy
$E(2^-) = 0.842$ MeV \cite{Cm246}, and in $^{248}$Cf with a much
lower bandhead $E(2^-) = 0.592$ MeV \cite{Cf248}. The recent data
have added new evidences for $^{250}$Fm with $E(2^-) = 0.879$ MeV
\cite{Fm250} and for $^{252}$No with $E(2^-) = 0.929$ MeV
\cite{No252}. We find it difficult to interpret these bands as
2-quasiparticle (qp) excitations in a reflection-symmetric well
because of their low bandhead energy, particularly for $^{248}$Cf.
The reason is simple. To be excited, a 2-qp state must have an
energy greater than two times of the BCS pairing gap, and the
pairing gaps in this mass region are typically $\sim$ 0.5 MeV for
neutrons and $\sim$ 0.8 MeV for protons, which are the needed amount
to correctly reproduce the rotational behavior, e.g. the moment of
inertia. Therefore, any (relatively) pure 2-qp state should lie
higher than 1 MeV above the ground state. For the observed
low-energy $2^-$ bands, there must be strong correlations that
sensitively act on the negative-parity configurations to push them
down. The primary candidate of the correlations is of the octupole
type.

The purpose of the present article is to investigate the role of
the nonaxial-octupole correlations in collective excitations of
transfermium and superheavy nuclei. We first analyze the
systematical trend of experimental low-lying spectroscopy data,
associated with axial-octupole and nonaxial-octupole modes and
ranging from the $A=220$ (actinide) to the 250 (transfermium) mass
region. This analysis convincingly shows the importance of
introducing nonaxial-octupole correlations in the description of
low-lying $2^-$ bands in the $A=250$ mass region. Second, we
generalize the Reflection Asymmetric Shell Model (RASM)
\cite{RASM} to include the triaxial-octupole degree of freedom,
which is applied to describe quantitatively the 2$^-$ bands. Based
on these results, we interpret these bands in the heaviest nuclear
systems as having connection to the predicted tetrahedral
symmetry.

\begin{figure}
\includegraphics[angle=-90,width=8cm]{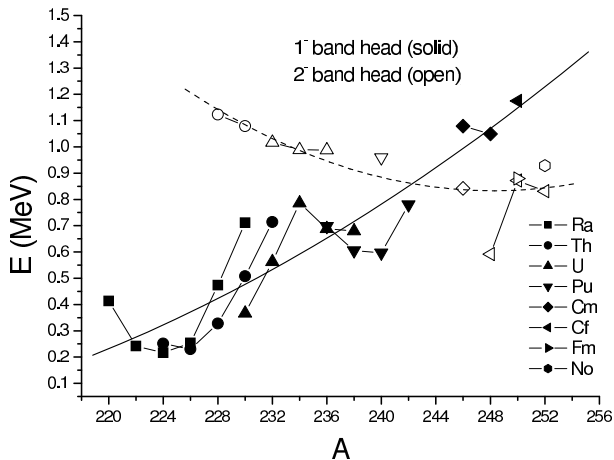}
\caption{Experimental energies of the 1$^-$ (solid symbols) and
2$^-$ (open symbols) bandheads of the isotopes Ra, Th, U, Pu, Cm,
Cf, Fm, and No. Data for Ra, Th, U, and Pu are taken from Refs.
\cite{Ra-U,Pu}, and for Cm, Cf, Fm, and No from Refs.
\cite{Cm246,Cf248,Fm250,No252}. The solid line and dash line are the
polynomial fits for the 1$^-$ and 2$^-$ bandhead energies,
respectively. Note that the two curves cross at $A=242$.}
\label{fig1}
\end{figure}

Experimental energies of the lowest 1$^-$ and lowest 2$^-$
bandheads of Ra, Th, U, Pu, Cm, Cf, Fm, and No are plotted as
functions of mass number in Fig. 1. To qualify a ``band", we use a
criterion which requires that for each band, at least three lowest
states be assigned experimentally. As a general tendency, it is
seen from Fig. 1 that the 1$^-$ bandhead energies increase with
increasing mass number, while the 2$^-$ ones behave like a
gradually decreasing function of mass number. In the Ra-Pu region,
the 1$^-$ bands lie very low in energy while the 2$^-$ bands are
quite high. This reflects the well-known fact that in Ra-Pu
nuclei, the axial-octupole correlation, associated with the
Y$_{30}$ shape, is important. However, the situation becomes
opposite in the Cm-Cf-Fm-No region where the 2$^-$ bandhead
energies drop to lower
values. 
The occurrence of low-lying 2$^-$ bands indicates that the
nonaxial-octupole effect becomes important in the transfermium
region.

The polynomial fits for the 1$^-$ (solid curve) and 2$^-$ (dash
curve) bandhead energies across each other at $A=242$, as seen in
Fig. 1. This suggests a spontaneous breaking of the axial-octupole
symmetry around this mass number. More precisely, the dominating
component of octupole correlation changes from the axial Y$_{30}$
at $A<242$ to the nonaxial Y$_{32}$ at $A>242$. In studies of the
nuclear reflection asymmetry in actinide nuclei with $A<242$, one
usually neglects, as a good approximation, the nonaxial-octupole
deformation and adopts only the axial Y$_{30}$ component in many
model calculations for the low-lying alternative parity bands
(see, for example, Ref. \cite{LC1988}). Following the same
philosophy, one should apply the nonaxial-octupole Y$_{32}$
deformed mean field as a useful starting point to model the
low-lying $2^-$ bands in transfermium and superheavy nuclei, where
the axial-octupole Y$_{30}$ deformation component can be neglected
as a good approximation. Of course, the quadrupole deformation
that is superposed with octupole deformation must be included to
describe the experimental moments of inertia in both actinide and
transfermium nuclei.

\begin{figure}
\includegraphics[width=8cm]{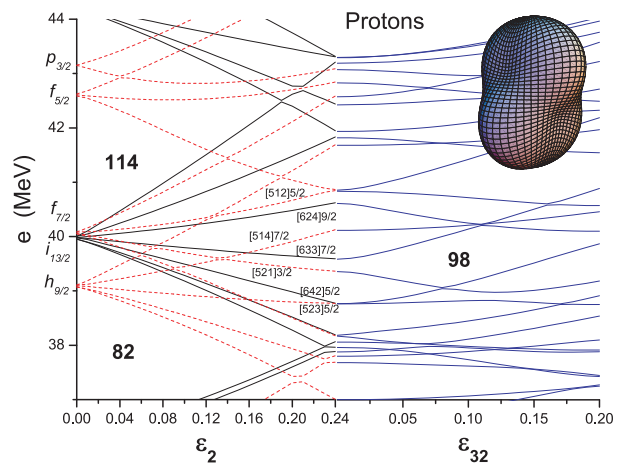}
\caption{(Color online) Proton Nilsson diagram showing
single-particle orbitals as deformation $\varepsilon_2$ varies (left
part), and as functions of $\varepsilon_{32}$ with fixed
$\varepsilon_2=0.24$ (right part). In the left part, orbitals with
positive (negative) parity are shown by solid (dashed) curves. }
\label{fig2}
\end{figure}

In an effective treatment with separable forces, the octupole effect
is described by the long-range octupole-octupole ($O$-$O$)
interactions between nucleons, which are produced by the matrix
elements of $Y_{3\mu}$ between single-particle states with $\Delta j
= \Delta l =3$ ($j$ and $l$ are the total angular momentum and orbit
angular momentum of single-particles, respectively). The condition
is found satisfied in several mass regions in the nuclear chart, and
particularly in those transitional regions between spherical and
well-deformed shapes. Thus, if the Fermi surface of a nucleus lies
close to a pair of orbitals with $\Delta j = \Delta l =3$, the
maximum octupole effect is expected. To see whether octupole
correlations are present in the heavy mass region, we plot in Fig. 2
the deformed single-particle states calculated with the Nilsson
potential. We note that this proton single-particle diagram created
by using the Nilsson parameters of Bengtsson and Ragnarsson
\cite{BR85} is very similar to the diagram calculated with the
Wood-Saxon potential \cite{Chas77}. The latter diagram has recently
been used to assign experimental band structure in the odd-proton
nucleus $^{249}$Bk \cite{Bk249}. In the left part of Fig. 2, proton
Nilsson diagram is plotted as a function of the quadrupole
deformation $\varepsilon_2$. The calculation predicts a sizable
spherical energy gap at proton number 114. It is particularly
notable that below the gap, the positive-parity state $i_{13/2}$
(with $l=6$ and $j=13/2$) and the negative-parity one $f_{7/2}$
(with $l=3$ and $j=7/2$) appear to be nearly degenerate at the
spherical limit. These two nearly-degenerate states fulfil precisely
the $\Delta j = \Delta l=3$ condition for the octupole correlation.
We note that the near-degeneracy of the proton $i_{13/2}$ and
$f_{7/2}$ states is also a result of the Wood-Saxon single-particle
diagram \cite{Chas77}.

A careful inspection on the left part of Fig. 2 leads to another
intersting discovery: When going away from the spherical limit and
moving to the deformed region, one can clearly see the pairs of
orbitals with opposite parities that are very close in energy and
stay nearly parallel as $\varepsilon_2$ varies. These are the pairs
$\{ [633]7/2, [521]3/2\}$ and $\{ [624]9/2, [512]5/2\}$. The two
orbitals in each pair satisfy the $\Delta j = \Delta l=3$ condition,
and in addition, are governed by $\Delta K=2$ ($K$ is the projection
of $j$ on the symmetry axis of a deformed nucleus). Thus these
orbitals are connected by large matrix elements of $Y_{32}$,
implying a strong triaxial-octupole correlation. The nuclei, whose
proton Fermi surfaces fall in the proximity of these orbitals, may
exhibit an $\varepsilon_{32}$ shape. The inserted figure at the
upper-right corner of Fig.2 illustrates the most favorable shape of
a nucleus with triaxial-octupole deformation superposed with
quadrupole deformation. Due to the strong octupole correlation the
collective negative-parity states (in this case the sequence with
the bandhead $I^\pi=2^-$) can be pushed down significantly.

It may possibly be true that the strong $Y_{32}$ correlation, with
the occurrence of very low-lying $2^-$ bands, is a special property
in superheavy nuclei. The strict conditions for this to happen are
the presence of nearly-degenerate pair(s) of single-particle states
at $\varepsilon_2=0$ that differ in the quantum numbers $l$ and $j$
by 3, and due to this degeneracy at the spherical limit, the
presence of closely-lying, opposite-parity, sub-orbitals for
$\varepsilon_2 > 0$ that differ in the quantum number $K$ by 2.

In the right part of Fig. 2, the diagram is obtained by including
an additional $Y_{32}$ term in the Nilsson Hamiltonian varying
with the octupole deformation parameter $\varepsilon_{32}$, while
the quadrupole deformation is kept at $\varepsilon_2=0.24$. As
expected, the two $\Delta K=2$ pairs feel a strong $Y_{32}$
interaction so that the force causes large energy splittings in
the diagram. As a consequence, an energy gap at proton number 98,
and a one at 106, show up. This result implies that nuclei in the
mass-250 region, with proton number from 96 to 108, will possibly
exhibit a strong triaxial-octupole effect with the Y$_{32}$-shape.
Particularly, isotopes of the element 98 (Californium) and 106
(Seaborgium) can feel very strong nonaxial-octupole correlation.
However, much less Y$_{32}$ effect is observed around N=150 in the
neutron single-particle diagram.

It will be more convincing if the above arguments at the mean-field
level are supported by some quantitative calculations. If the
observed $2^-$ bands are indeed a manifestation of the
triaxial-octupole correlation, the calculations that properly
include the octupole degree of freedom should be able to reproduce
them. To demonstrate this, we calculate the low-lying spectrum in
these nuclei by using the RASM \cite{RASM}. The RASM extends the
Projected Shell Model \cite{PSM} applicability by breaking the
reflection symmetry in the basis and carrying out parity projection
in addition to angular momentum projection. In this way, the
octupole correlation is efficiently described and the model is thus
particularly suitable to be applied to nuclei that show octupole
effects. The trail wave function of the triaxial RASM is written in
terms of parity- and angular-momentum-projected multi-qp states
\begin{equation}
|\Psi_{IpM}\rangle = \sum_{K\kappa} f^{Ip}_{K\kappa} \hat P^p \hat
P^{\,I}_{MK_\kappa}|\phi_\kappa\rangle , \label{wf}
\end{equation}
where the index $\kappa$ labels deformed basis states, and $\hat
P^p$ and $\hat P^{\,I}_{MK}$ are the parity-projection operator and
angular-momentum-projection operator, respectively. For even-even
nuclei, $|\phi_\kappa\rangle$ in Eq. (\ref{wf}) is
\begin{equation}
\{ |0\rangle, a^\dagger_\nu a^\dagger_\nu |0\rangle, a^\dagger_\pi
a^\dagger_\pi |0\rangle, \cdots \} \label{space}
\end{equation}
where $a^\dagger_\nu$ and $a^\dagger_\pi$ are the qp creation
operator for neutrons and protons, respectively, with respect to the
qp vacuum $|0\rangle$. The qp states are obtained from a deformed
Nilsson calculation followed by a BCS calculation, in a model space
with three major shells for each kind of nucleon (major shells 5, 6,
7 for neutrons and 4, 5, 6 for protons). The deformed Nilsson
calculation is performed with the quadrupole term and the octupole
term; The former term breaks the rotational symmetry, and the later
the reflection symmetry, in the qp states. The utility of symmetry
breaking is to introduce efficiently the corresponding correlations
in the basis states. The symmetries thus spontaneously broken are
recovered in the total wave function by projection (see Eq.
(\ref{wf})).

The two-body Hamiltonian has the form
\begin{equation}
\hat H = \hat H_0 - {1\over 2} \sum_{\lambda = 2}^3 \chi_\lambda
\sum_{\mu = -\lambda}^\lambda \hat Q_{\lambda\mu}^\dagger \hat
Q_{\lambda\mu} - G_0 \hat P_{00}^\dagger \hat P_{00} - G_2 \sum_{\mu
= -2}^2 \hat P_{2\mu}^\dagger \hat P_{2\mu}. \label{hamilt}
\end{equation}
In Eq. (\ref{hamilt}), $\hat H_0$ is the spherical single particle
Hamiltonian, the second term contains the quadrupole-quadrupole and
octupole-octupole interactions, and the third and the last term are
the monopole pairing and quadrupole pairing force, respectively. The
interaction strengths in the second term are related to the
deformation parameters through self-consistent relations
\begin{eqnarray}
\chi_{2,\tau\tau^\prime} &=& \sqrt{{4\pi}\over 5} {{{2\over
3}\varepsilon_2 \hbar \omega_\tau \hbar \omega_{\tau^\prime}}\over
{\hbar \omega_n \langle Q_{20}\rangle_n + \hbar \omega_p \langle
Q_{20}\rangle_p}}, \\
\chi_{3,\tau\tau^\prime} &=& \sqrt{{\pi}\over 7} {{\varepsilon_{32}
\hbar \omega_\tau \hbar \omega_{\tau^\prime}}\over {\hbar \omega_n
\langle Q_{32}\rangle_n + \hbar \omega_p \langle Q_{32}\rangle_p}}.
\label{chi}
\end{eqnarray}
Therefore, for given deformations, $\chi$ are calculated, and thus
not regarded as parameters. In the present work, we fix
$\varepsilon_2=0.235$ for all nuclei; this value is verified to be
consistent with that obtained from the TRS calculations. The
octupole deformation parameter $\varepsilon_{32}$ influences
sensitively the energy separation of the $2^-$ band from the ground
state. We find that to reproduce data, the best suitable values are:
$\varepsilon_{32}=$ 0.105 for $^{246}$Cm, 0.118 for $^{248}$Cf,
0.110 for $^{250}$Fm, and 0.107 for $^{252}$No. We note that
$^{248}$Cf with $Z=98$ requires the largest $\varepsilon_{32}$ among
the isotones. The monopole pairing strength $G_0$ takes the usual
form
\begin{equation}
G_0 = {{g_1\mp g_2{{N-Z}\over A}}\over A}, \label{G0}
\end{equation}
where the minus (plus) sign stands for neutrons (protons), and
$g_1=20.36$ and $g_2=11.26$ are fixed for all the nuclei studied in
this paper. The quadrupole pairing strength $G_2$ is assumed to be
proportional to $G_0$, with the proportional constant 0.13.

\begin{figure*}
\includegraphics[angle=-90,width=14cm]{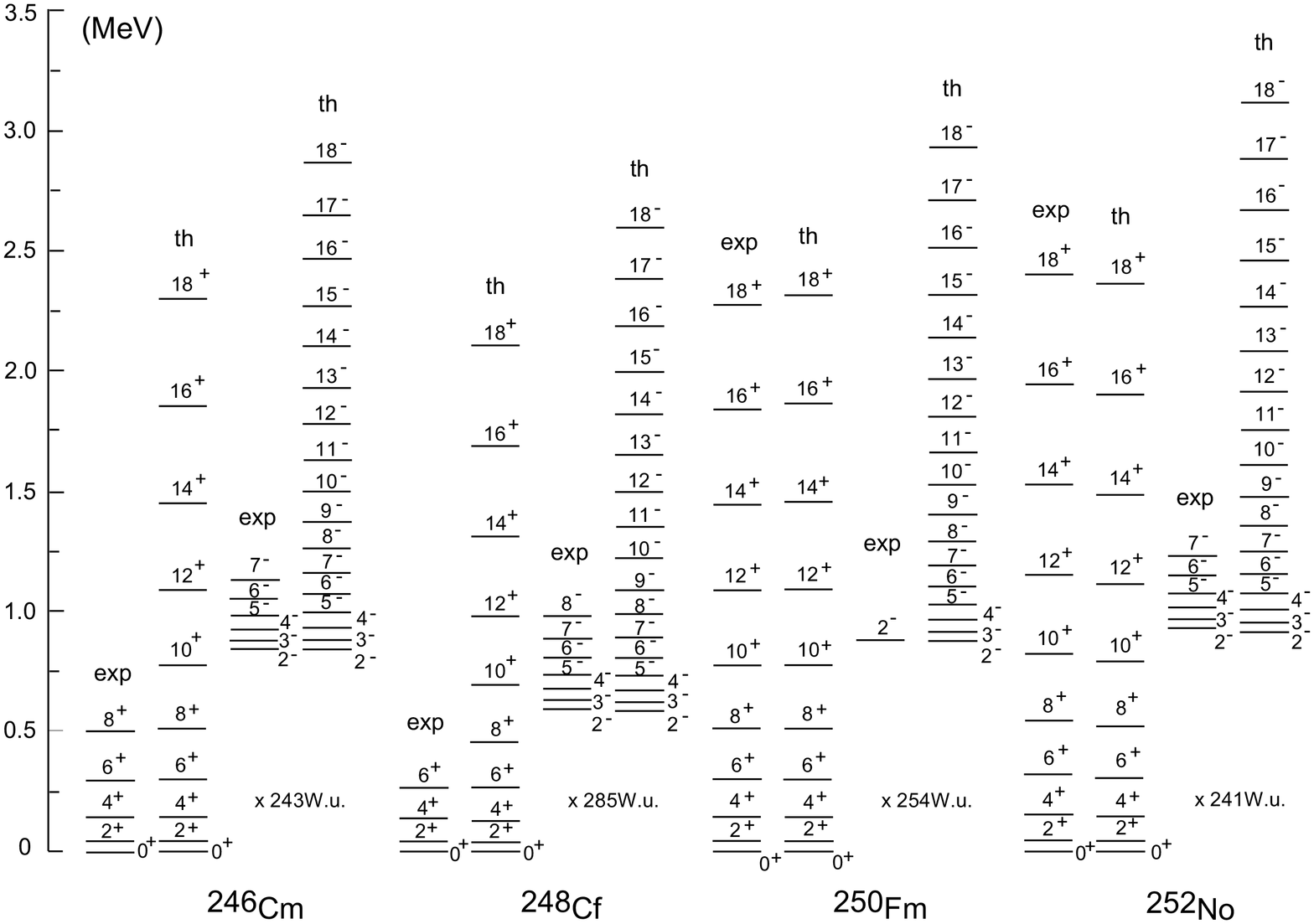}
\caption{The RASM calculations for the ground band and $2^-$ band in
the $N=150$ isotones. The theoretical results are compared with
available data. Note the very low $2^-$ band in $^{248}$Cf. The
calculated B(E3; $0^+\rightarrow3^-$) value in Weisskopf unit is
marked by `x'. } \label{fig3}
\end{figure*}

Diagonalizing the Hamiltonian (\ref{hamilt}) in the space spanned by
parity- and angular-momentum-projected multi-qp configurations
(\ref{space}), we obtain a set of energy levels. From them, we plot
in Fig. 3 the ground band (the lowest $\Delta I=2$ rotational
sequence with positive-parity) and the $2^-$ band (the $\Delta I=1$
sequence with negative-parity), both of which are compared with
available data. As one can see in Fig. 3, agreement between theory
and experiment is excellent. We also obtain a good agreement with
the unpublished $2^-$ band in $^{250}$Fm \cite{GreenP}. A careful
look indicates that, to compare with the neighboring isotones, the
nucleus $^{248}$Cf has a notably low $2^-$ band. It is remarkable
that this low $2^-$ band can be well reproduced by using an enhanced
$\varepsilon_{32}$ in the calculation (see the employed
$\varepsilon_{32}$ values in the last paragraph), reflecting the
strongest triaxial-octupole correlation in this nucleus. The physics
behind can be understood by looking at the right part of Fig. 2. It
can be clearly seen that there is an energy gap at the proton number
98. The emergence of a gap is a consequence of the maximum
triaxial-octupole interaction, and therefore, $^{248}$Cf shows the
lowest $2^-$ band. For the same reason, we expect very low energy
$2^-$ band in the Sg nuclei (the element 106).

These low-lying 2$^-$ bands are collective in nature. The degree of
octupole collectivity is generally determined by an enhancement of
reduced $E3$ transition probability $B(E3)$ over the single-particle
estimate.  The reduced rate of electromagnatic transitions between
the eigenstates $\left|\psi_{IpM}\right>$, induced by a spherical
tensor operator $\hat M_{\lambda\mu}$, is
\begin{equation}
B(M\lambda;i\rightarrow f)=\frac{2I_f+1}{2I_i+1}\left|\left<
\Psi^{\sigma_f}_{I_f\pi_f}\|\hat
M_{\lambda}\|\Psi^{\sigma_i}_{I_i\pi_i}\right>\right|^2,
\label{reducetrans}
\end{equation}
and the reduced matrix elements become
\begin{eqnarray}
&& \left< \Psi^{\sigma_f}_{I_f\pi_f}\|\hat
M_{\lambda}\|\Psi^{\sigma_i}_{I_i\pi_i}\right>=
\frac{1}{2}\left(1+\pi_f\pi_i\pi_\lambda\right) \nonumber\\
 && \sum_{K_i\kappa_iK_f\kappa_f} f^{\sigma_fI_f\pi_f}_{K_f\kappa_f}
f^{\sigma_iI_i\pi_i}_{K_i\kappa_i}
\sum_\mu\left<I_iK_f-\mu\lambda\mu|I_fK_f\right> \nonumber \\
&&\left<\Phi_{\kappa_f}\left|\hat
M_{\lambda\mu}P^{I_i\pi_i}_{K_f-\mu
K_i}\right|\Phi_{\kappa_i}\right>. \label{reducematrix}
\end{eqnarray}
The calculated reduced transition probabilities from ground state
$0^+$ to the $3^-$ state of the $2^-$ band, $B(E3; 0^+\rightarrow
3^-)$, are shown in Fig. 3 in Weisskopf (single-particle) units
(W.u.) (see the values marked by `x'). Again, the most enhanced
triaxial-octupole collectivity appears in $^{248}$Cf, where the
$B(E3)$ value is the largest.

As we are exploring a completely unknown island of superheavy
elements, it is very useful to understand the structure of the
mass-250 nuclei, which are the gateway to the superheavy region.
Most studies on the excited configurations to date have been
focused on the multi-quasiparticle structure (see, for example,
Refs. \cite{Rodi06,Tandel06}). Recently, alternative parity bands
in the heaviest nuclei are predicted \cite{Shnei06}. The excited
configurations are sensitive to the detailed shell structure, and
thus can bring more useful information. The degeneracy of the
proton $i_{13/2}$ and $f_{7/2}$ states found in the
single-particle diagram may be accidental; however, it creates the
right condition to form the strong triaxial-octupole correlation
that explains the experimental data. Thus, the present proposal
seems to set up a new, strong constraint on all mean field
calculations that are employed to predict shell structure for the
superheavy mass region.

In conclusion, the nonaxial-octupole $Y_{32}$ correlation has been
found to be important in the transfermium and superheavy region
for the elements from 96 to 108, with the strongest effect in the
elements 98 and 106. We have demonstrated that if this correlation
is considered in realistic RASM calculations, the experimentally
observed low-energy 2$^-$ bands in the $N=150$ isotonic chain can
be easily described. The present work has thus suggested that this
exotic nuclear effect, which has been discussed theoretically in
other mass regions but with no convincing experimental evidences,
may show up in the heaviest nuclei. The unique role of the
nonaxial-octupole degrees of freedom gives weight to the recent
prediction of tetrahedral symmetry in atomic nuclei. It has also
set up a constraint on theoretical single-particle states for the
superheavy region. The reduced E3 transition probabilities for the
Coulomb excitation from the $0^+$ ground states to the $3^-$
states of the $2^-$ bands are predicted to be strong. The future
measurements of E3 transition rates would be crucial for the final
determination of the nonaxial octuple nature of the $2^-$ bands
and, consequently, their relation to the tetrahedral symmetry.

The authors thank R.-D. Herzberg and P. T. Greenlees for sharing the
data prior to publication, and P. M. Walker for stimulating
discussions. This work is supported by NNSF of China under contract
No. 10475115, 10305019, 10435010, by the Chinese Major State Basic
Research Development Program through grant 2007CB815005, and by the
the U. S. National Science Foundation through grant PHY-0216783.



\end{document}